\documentclass[preprint]{aastex}
\shorttitle{Interaction of two filaments}
\slugcomment{Submitted to ApJ}
\shortauthors{Zheng et al.}
\begin{document}

\title{Interaction of two filaments in a long filament channel associated with twin coronal mass ejections}
\author{Ruisheng Zheng$^{1}$, Qingmin Zhang$^{2}$, Yao Chen$^{1}$, Bing Wang$^{1}$, Guohui Du$^{1}$, Chuanyang Li$^{1}$, and Kai Yang$^{3}$}
\affil{$^{1}$Shandong Provincial Key Laboratory of Optical Astronomy and Solar-Terrestrial Environment, and Institute of Space Sciences, Shandong University, 264209 Weihai, China; ruishengzheng@sdu.edu.cn\\
 $^{2}$Key Laboratory for Dark Matter and Space Science, Purple Mountain Observatory, CAS, Nanjing 210008, China\\
 $^{3}$School of Astronomy and Space Science, Nanjing University, 210023 Nanjing, China\\}

\begin{abstract}
Using the high-quality observations of the {\it Solar Dynamics Observatory}, we present the interaction of two filaments  (F1 and F2) in a long filament channel associated with twin coronal mass ejections (CMEs) on 2016 January 26. Before the eruption, a sequence of rapid cancellation and emergence of the magnetic flux has been observed, which likely triggered the ascending of the west filament (F1). The east footpoints of rising F1 moved toward the east far end of the filament channel, accompanying with post-eruption loops and flare ribbons. It likely indicated a large-scale eruption involving the long filament channel, resulted from the interaction between F1 and the east filament (F2). Some bright plasma flew over F2, and F2 stayed at rest during the eruption, likely due to the confinement of its overlying lower magnetic field. Interestingly, the impulsive F1 pushed its overlying magnetic arcades to form the first CME, and F1 finally evolved into the second CME after the collision with the nearby coronal hole. We suggest that the interaction of F1 and the overlying magnetic field of F2 led to the merging reconnection that form a longer eruptive filament loop. Our results also provide a possible picture of the origin of twin CMEs, and show the large-scale magnetic topology of the coronal hole is important for the eventual propagation direction of CMEs.
\end{abstract}


\section{Introduction}
Coronal mass ejections (CMEs), the most spectacular phenomena occurring on the Sun, often consist of abundant structures comprising plasma and magnetic fields that are expelled from the Sun into {\bf the} interplanetary space. If two CMEs erupt closely in time from the same active region (AR), they are called as  ``twin-CMEs". According to Gopalswamy et al. (2004) and Li et al. (2012), this kind of ``twin-CME" event may facilitate the particle acceleration in the solar atmosphere and in the interplanetary space. It is suggested that if the preceding CME drives a shock and creates an environment of enhanced level of turbulence and energetic particles, the following CME with its associated separate shock may propagate into this environment, leading to a more efficient shock acceleration process (see also, Li \& Zank 2005; Ding et al. 2013, 2014). Hence, twin CMEs, representative of a special agent capable of efficient particle acceleration, are crucial in both solar physics studies and space weather forecasting. However, the physical origin and eruption mechanism of twin CMEs remain elusive.

It is widely accepted that filament eruptions are closely associated with solar flares and CMEs (Low 1996; Chen 2011; Schmieder et al. 2013), and the three eruptive phenomena are different manifestations of a single process (e.g., Zhang et al. 2001). There are many models for the eruption mechanisms, such as the tether-cutting reconnection (Moore et al. 2001), the breakout reconnection (Antiochos et al. 1999; Chen et al. 2016), the helical kink instability (Fan 2005), and the torus instability (Kliem \& T\"or\"ok 2006; Aulanier et al. 2010). In the scenario of the common pattern of filament eruptions (Sterling \& Moore 2004), the filament gradually ascends in an initial slow-rise phase, followed by a sharp change to a fast-acceleration phase. In addition to successful eruptions, there also exist failed/confined eruptions (Ji et al. 2003; Liu et al. 2009; Guo et al. 2010; Joshi et al. 2013; Song et al. 2014; Dalmasse et al. 2014), and partial eruptions (Gilbert et al. 2007; Liu et al. 2007; Shen et al. 2012; Zhang et al. 2015).

Based on three-dimensional magnetohydrodynamic simulations for twisted flux tubes under convective zone conditions, Linton et al. (2001) found four fundamental types of flux-tube interactions: bounce, merging, slingshot, and tunnel. They also proposed that the merging interaction has significant reconnection involving mostly the azimuthal fields, and the two interacting flux tubes merge into one after the interaction. Filaments are believed to embed in the magnetic configurations of filament channels or flux ropes (Aulanier et al. 2002; Liu et al. 2012), and the merging interactions can occur between filaments along one common filament channel or along different filament channels (Martin et al. 1994; Schmieder et al. 2004; DeVore et al. 2005; Aulanier et al. 2006; Su et al. 2007; Bone et al. 2009; Chandra et al. 2011; Jiang et al. 2014; Joshi et al. 2014, 2016). The newly-formed filament can eventually erupt, if it is destabilized by further magnetic cancellation at the footpoints (Litvinenko \& Martin 1999; Martens \& Zwaan 2001; DeVore et al. 2005).

In this study, we present the observations of the interaction between two filaments in a same filament channel associated with twin CMEs, by using high-quality data from the Atmospheric Imaging Assembly (AIA: Lemen et al. 2012) onboard the {\it Solar Dynamics Observatory (SDO}: Pesnell et al. 2012). The paper is organised as follows. Section 2 describe the observations used in the work; The main results are presented in Section 3; We give the conclusions and discussion in Section 4.

\section{Observations and Data Analysis}
The filament eruption occurred at the NOAA AR 12486 on 2016 January 26, accompanying with a C1.3 flare. To study the filament evolution, we utilise the observations from SDO/AIA combining the H$\alpha$ filtergrams from the Global Oscillation Network Group (GONG) of the National Solar Observatory. The AIA observes the full disk (4096~$\times$ 4096 pixels) of the Sun and up to 0.5 $R_{sun}$ above the limb in ten EUV and UV wavelengths, with a pixel resolution of 0$\farcs$6 and a cadence of 12 s. The H$\alpha$ images are at 6563~{\AA} with a spatial resolution of 1$\arcsec$ and a cadence of $\sim$1 minute (Harvey et al. 2011). Magnetograms from the Helioseismic and Magnetic Imager (HMI: Scherrer et al. 2012), with a cadence of 45 s and pixel scale of 0$\farcs$6, are used to check the magnetic field evolution of the source region. The evolution of CMEs in the high corona is detected by the Large Angle and Spectrometric Coronagraph (LASCO; Brueckner et al. 1995) C2. The kinematics of the filament and associated moving plasma are analyzed with the time-slice approach. The speeds are determined by linear fits, with the measurement uncertainty taken to be 4 pixel ($\sim$1.74 Mm) for AIA data. We also use the Potential Field Source Surface (PFSS; Schrijver {\&} De Rosa 2003) model to extrapolate the large-scale magnetic field topology in the corona.

\section{Results}
\subsection{Magnetic Activities}
Figure 1 shows the magnetic filed evolution of AR 12486 in HMI magnetograms. The decaying AR consists of a leading positive polarity and a following negative polarity. The magnetic polarities are dispersed, and the long polarities inversion line (PIL) has less shear. Most of the opposite magnetic polarities along the PIL are disconnected. The polarities along the southern part of the PIL are diffuse and weak, and that along the northern part of the PIL are closer and stronger (panel a). There exists obvious cancellation and emergence of magnetic flux along the PIL (see the attached animation), and we focus on the magnetic field on both sides of the northern part of the PIL. There are some magnetic cancellation sites, where the opposite polarities are approaching and colliding with each other (white boxes in panels b-h). Between the unconnected opposite magnetic polarities, some magnetic flux emerged in the form of small patches or spots (cyan arrows in panels d-f), followed by quick cancellations. In the plot of the total flux change (panel i) for the magnetic field in the northern of AR (the box in panel a), for both the positive and negative fluxes, there has a trend of decreasing since $\sim$08:00 UT (the left vertical line), and it is very clear for the abrupt cancellation and rapid emergence just before the eruption (the right vertical line and arrows in panel i).

\subsection{Filament Eruption}
Figure 2 shows the filament evolution (see the attached animations) in composite warm passbands (AIA 171, 193, and 211~{\AA}; panels a, d, and g) and in AIA 304~{\AA} (panels b-c, e-f, and h-i). In 304~{\AA}, two filament systems, the west filament (F1) and the east filament (F2; white arrows), located in the same long filament channel. F1 and its overlying magnetic arcades are shown in details in the right-down corner, and the cool material of magnetic arcades condensed in shallow dips along magnetic field lines (panels a-b). Note that a polar coronal hole (CH; yellow arrows) is southeast to the AR. Likely due to the magnetic activities, the filament system {\bf was} activated, and there appeared some brightenings around the east footpoints (B1 and B2) of F1 and the magnetic arcades, which made F1 and the magnetic arcades more obvious. On the other hand, the dark F2 was clearly seen to underlie F1 and magnetic arcades, and B1-B2 both located closely in the north to the western part of F2 (white arrows in panels c-e). After the activation, F1 and the magnetic arcades rose slowly, and F1 was heated to become brighter than the cooler magnetic arcades (panel f). At $\sim$17:25 UT, the heated F1 began to erupt southward, and two flare ribbons appeared at the center of the filament channel (black arrows in panels h-i). Interestingly, the brightened B1 moved southeastward along the filament channel, while B2 was nearly fixed during the eruption (B1 and B2 in panels c-i). Moreover, it seemed to be disconnected between the top of F1 and B1 (green arrows) during the eruption, and some bright plasma (blue arrows) moved from the intersection of F1 and F2 toward the far east end of the filament channel (panels h-i).

Figure 3 displays the filament activities in H$\alpha$ filtergrams. Superposed with the HMI magnetogram, F1 and F2 were clearly located along the PIL, and the filament center was wide (panel a). Before the eruption, F1 and its magnetic arcades were already highly activated, which was especially obvious in the middle wide part (see the attached animation). At $\sim$16:34 UT, B1 clearly coincided with the west footpoints of F2 (the southwestward green arrow), and some portions of F2 were undetectable, due to the low opacity (panel b). After the activation, F1 was heated in a higher temperature and became invisible in H$\alpha$, and the magnetic arcades (pink arrows) began to lift up, resulted in some enhanced brightenings (red arrows) around their west footpoints, while F2 (white arrows) became thin and faint (panels c-d). At $\sim$17:22 UT, the magnetic arcades began to be disrupted with cracks in their tops (northward pink arrows in panels e-h), as a result of being pushed by the erupting F1. During the eruption, B2 barely moved, but B1 slipped southeastward (B1 and B2 in panels e-i). As a result of the eruption, flare ribbons (black arrows) and some mass flow (blue arrows) are also distinct (panels f-i). The bright plasma came from the intersection of F1 and F2, and became dark during its eastward motion over F2 (panel i). Followed with the mass flow, the western part of F2 disappeared, and the eastern portions of F2 were mostly at rest (white arrows in panels h-i).

\subsection{Interaction of Filaments}
The erupting F1 and the expanding magnetic arcades are clearly seen in other AIA passbands (panels a-b of Figure 4). In addition, it is also clear for the moving plasma extending eastward to the far end of the filament channel (blue arrows), and the post-eruption loops (the yellow arrow) appeared earlier in AIA hot passbands than that in warm passbands (panels d-e). The evolution of the eruption is better seen in AIA 171 and in 304~{\AA} running-difference images (panels c and f-g). The impulsive F1 ran fast southeastward, and magnetic arcades were pushed to ascend, with overlying loops expanding as a faint rim (white arrows in panel c). F1 kept moving southeastward, and separated clearly from magnetic arcades (arrows in panels f-g). Note that the erupting F1 seems to connect with the east end of the long filament channel. The evolution in two different directions (S1 and S2 in panel f) is clearly shown in time-slice plots (panels g-h). It is obvious that F1 and magnetic arcades both rose slowly in a few tens km s$^{-1}$ (black arrows) in the pre-eruption phase from $\sim$17:25 UT (dashed vertical lines). Along southwestward S1, magnetic arcades impulsively expanded at a speed of $\sim$139.7 km s$^{-1}$ (pink arrow), and F1 rose very slowly (green arrow). Along southeastward S2, the expanding magnetic arcades was faint, but F1 began to accelerate after the eruption and finally ceased at the boundary of the CH at $\sim$17:50 UT (the red arrow). Before its halt, F1 first accelerated to $\sim$108.6 km s$^{-1}$, and then obtained a faster speed of $\sim$305.6 km s$^{-1}$ (the green arrows).

Figure 5 shows the evolution of the three eruptive parts in AIA 304 and 171~{\AA} images (left and middle panels), and in time-slice plots (right panels). The rising magnetic arcades (the pink arrow) had a westward drift from 18:10 to 18:50 UT (panels a-b; see the attached animation), with a speed of $\sim$211.5 km s$^{-1}$ (panel c). After the encounter with the CH, the impulsive F1 was forced to move along the open field line of the CH, in the form of the moving bright feature (the green arrow in panel e). The F1 has a velocity of $\sim$179.5 km s$^{-1}$ (panel f). Note that there appeared brightenings and dimmings at the footpoints of the CH (cyan arrows in panels a-b and d-f), which likely indicates the interaction between the erupting F1 and the CH. Moreover, the moving plasma (the blue arrow) induced a longitudinal oscillation of F2 (the white arrow) in the same filament channel (panel g). In the time-slice plot (panel i), the oscillation clearly started after the arrival of the moving plasma that had an injection speed of $\sim$165.1 km s$^{-1}$. The initial speed of the oscillation is $\sim$47.3 km s$^{-1}$ (the red arrow), and the oscillation period is $\sim$55 minutes, estimated by the first two peaks (dotted lines), which is similar to the results of Zhang et al. (2012).

\subsection{Twin CMEs}
Following the eruption, there appeared a pair of (i.e., the twins) CMEs in the field of view (FOV) of LASCO C2 (Figure 6) {\footnote{\url{http://sidc.oma.be/cactus/catalog/LASCO/2\_5\_0/qkl/2016/01/latestCMEs.html}}}.
The first CME (CME1) appeared at 18:24 UT with a bright front (panel a), developing into a wider CME structure. The second CME (CME2) emerged later from the south end of CME1 with a narrow front (red arrow in panel b). The central position angles of the twin CMEs are 244$^\circ$ and 209$^\circ$, respectively, and their median speeds are 297 km s$^{-1}$ and 134 km s$^{-1}$, respectively. The close correlations in time, location, and velocity, demonstrate that the twin CMEs are closely associated with the filament eruption. It is likely that CME1 and CME2 came from magnetic arcades and F1, respectively. Both CMEs did not extend eastward during the expansion, due to the limit of the CH magnetic field. CME2 seemed to move along the south boundary of CME1 in a narrow extent, which is consistent with the moving bright feature along the open filed line in Figure 5.

After examining the PFSS results of the CH magnetic field lines (panel e), we see that the CH open field lines are of negative polarity lying next to the AR. The open field lines at the CH-AR boundary, with the non-radial expansion, occupy the space above the AR. This magnetic configuration helps understand the drift of rising magnetic arcades, and how the eruptive F1 ran into field lines of the CH, and then was forced to move outward in the form of CME2 along the specific trajectory.

\section{Conclusions and Discussion}
Here we study a very interesting filament eruption associate with two filaments (F1 and F2) in a same filament channel. The evolutionary details of the event are listed in Table 1. The ascent of F1 was likely triggered by a sequence of rapid cancellation and emergence, supported by their intimate spatial-temporal correlation. The interaction of F1 and F2 is evidenced by the east footpoints of F1 slipping toward the east far end of the filament channel and post-eruption loops straddling the intersection of two filaments. Due to the bright moving plasma and the longitudinal filament oscillation in the eastern part (involving F2) of the filament channel, we suggest that the interaction likely occurs between F1 and the part (higher) magnetic field overlying F2, which makes F2 nearly stayed at rest during the eruption.  The brightenings and dimmings at the CH boundary indicate the collision of the erupting F1 and the nearby CH. According to the close temporal-spatial correlation and the large-scale CH-AR magnetic field configuration given by PFSS, we suggest that the magnetic arcades were pushed to erupt as the first CME (CME1), and the erupting F1 finally evolved into the second CME (CME2) after the interaction with F2 and the collision with the CH.

\begin{deluxetable}{cccc}
\tabletypesize{\scriptsize}
\tablecaption{Timeline for the filament eruption on 2016 January 26}
\tablewidth{0pt}
\tablehead{Time &Observations}
\startdata
11:00-16:30 UT &F1 and magnetic arcades were destabilised, and  became wider. \\
16:34 UT &The east footpoints (B1) of F1 were very close to the west footpoints of F2. \\
17:00 UT &F1 and magnetic arcades started to lift slowly, and F2 was nearly at rest. \\
         &Brightenings appeared at B1 and B2.  \\
17:08-17:22 UT &B1 moved southeastward, and B2 was always fixed. \\
17:25 UT &F1 impulsively erupted southeastward, and pushed magnetic arcades to expand. \\
         &Flare ribbons and bright plasma flow were detected. \\
17:25-17:44 UT &The western part of F2 became invisible. \\
         &The eastern portions of F2 survived. \\
17:54 UT &The erupting F1 began to decelerate. \\
         &Some brightenings and dimmings appeared at the CH boundary. \\
18:00 UT &F1 propagated in a form of bright blob, and some dark material of magnetic arcades drifted westward. \\
        &There appeared a filament longitudinal oscillation in the filament channel of F2. \\
18:24 UT &The first CME (CME1) appeared in the FOV of LASCO/C2. \\
19:24 UT &The second CME (CME2) appeared in the FOV of LASCO/C2. \\[1ex]
\enddata

\end{deluxetable}

Based on the observations and the assumed magnetic filed overlying F2, we propose one possible interpretation in the schematic representation in Figure 7. Viewing F1 (green) and F2 (long cyan) from the positive-polarity side, their axial field points to the left (panel a). Hence, F1 and F2 both reveals sinistral chirality and the parallel axial fields, in accordance with the general hemispheric rule for the south hemisphere (Martin et al. 1994; Zirker et al. 1997; Pevtsov et al. 2003). Due to the southward motion, the axial field of the eastern part of F1 intersects the presumed higher field lines (orange lines) overlying F2 at an acute angle (black arrows in panels b-c), which is suitable for the merging reconnection (Linton et al. 2001). Following the continuous southward motion of the eruptive F1, the reconnection sites also shift along the axis of F2 at their contact points (panel c), and the east footpoints (B1) of F1 also slipped along the east footpoints (D1-D2) of higher field lines overlying F2. Finally, F1 and the part of F2 merge into a new longer filament structure (the red F in panels c-d), and magnetic arcades were pushed to expand with the fixed east footpoints (B2). Moreover, the merging reconnection likely only occurred between the erupting F1 and the overlying higher field lines of F2, and the overlying lower field lines (blue lines) made F2 be confined. Hence, the partial merging reconnection results in that most of F2 (short cyan in panels c-d) keep intact, and some bright plasma moves along the lower field lines (blue lines with white arrows) of F2. For the disappearance of the western part of F2 close to the flare ribbons, it is possible due to the mass transfer of the plasma flow or due to the heating process during the eruption. 

Intriguingly, the eruption involved a twin-CME eruption. In the twin-CME scenario proposed by Li et al. (2012), two CMEs originate from two eruptions, and take off closely in time from the same AR. Here, the twin CMEs formed due to the two erupting branches in different propagation directions. CME1 came from the expanding of overlying magnetic arcades driven by the impulsive F1/F. CME2 evolved from the impulsive F1 after the collision at the boundary of a nearby CH. The change of propagation direction of F1 and the westward material drift of the magnetic arcades (Fig.5a-c and the yellow arrow in Fig.7d) show the importance of large-scale magnetic topology of the CH to define the eventual propagating directions of eruptions (see also Zheng et al. 2016).



\acknowledgments

SDO is a mission of NASA's Living With a Star Program. The authors thank the SDO team for providing the data. This work is supported by grants NSBRSF 2012CB825601, NSFC 41274175, 41331068, 11303101, and 11603013, Shandong Province Natural Science Foundation ZR2016AQ16, and Young Scholars Program of Shandong University, Weihai, 2016WHWLJH07.

\clearpage

\begin{figure}
\epsscale{0.9}
\plotone{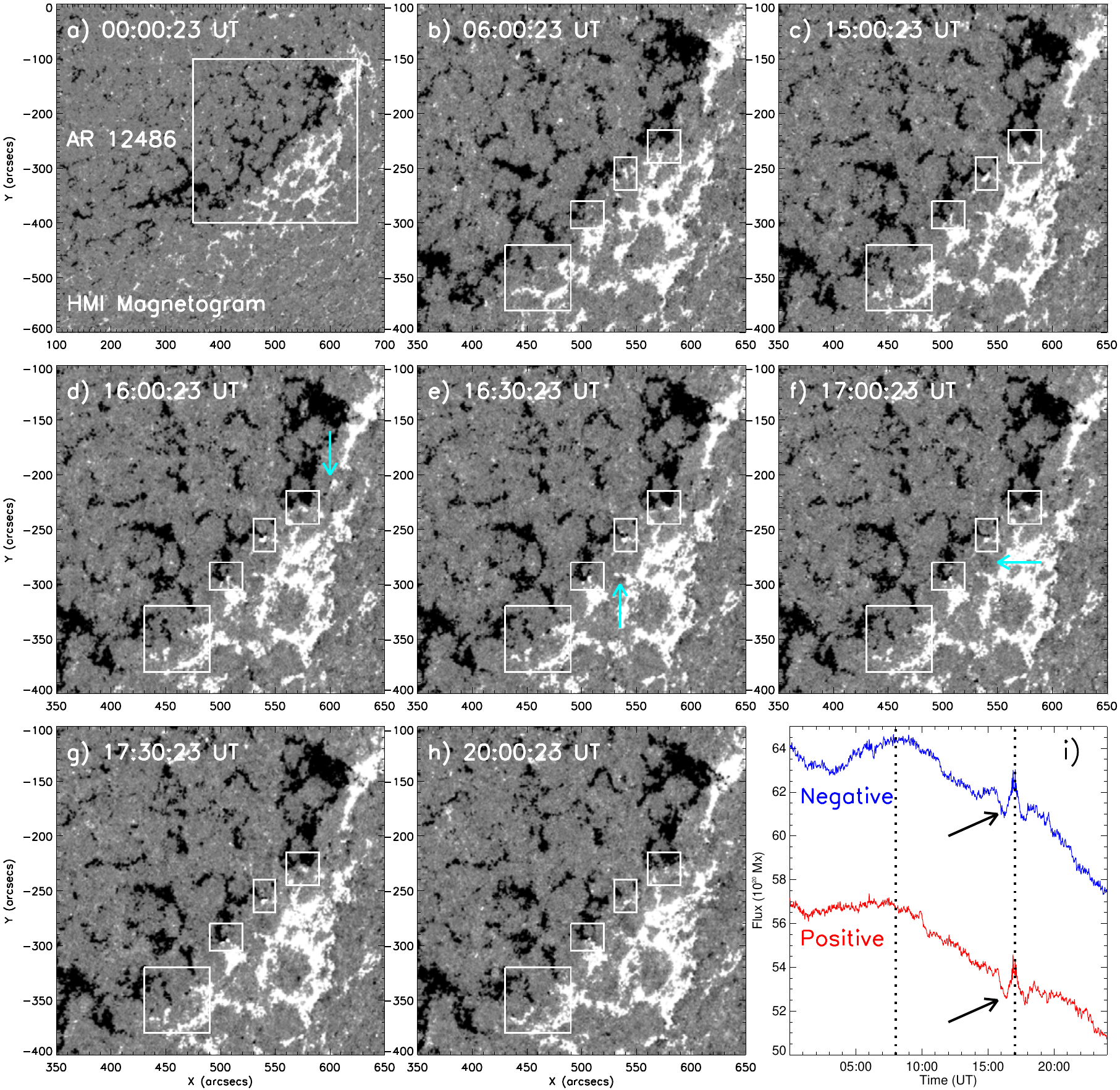}
\caption{(a)-(h) The magnetic filed evolution of AR 12486 in HMI magnetograms, (i) and the changes of the positive (red) and negative (blue) of the magnetic flux for the the box region that indicates the FOV of (b)-(h). In (b)-(h), the boxes show the magnetic cancellation regions, and cyan arrows point out the magnetic emergence sites. In (i), the arrows indicate the rapid cancellation and emergence before the eruption, and the left and right vertical lines mark the start of the cancellation and the eruption onset, respectively.
\label{f1}}
\end{figure}

\clearpage

\begin{figure}
\epsscale{0.9}
\plotone{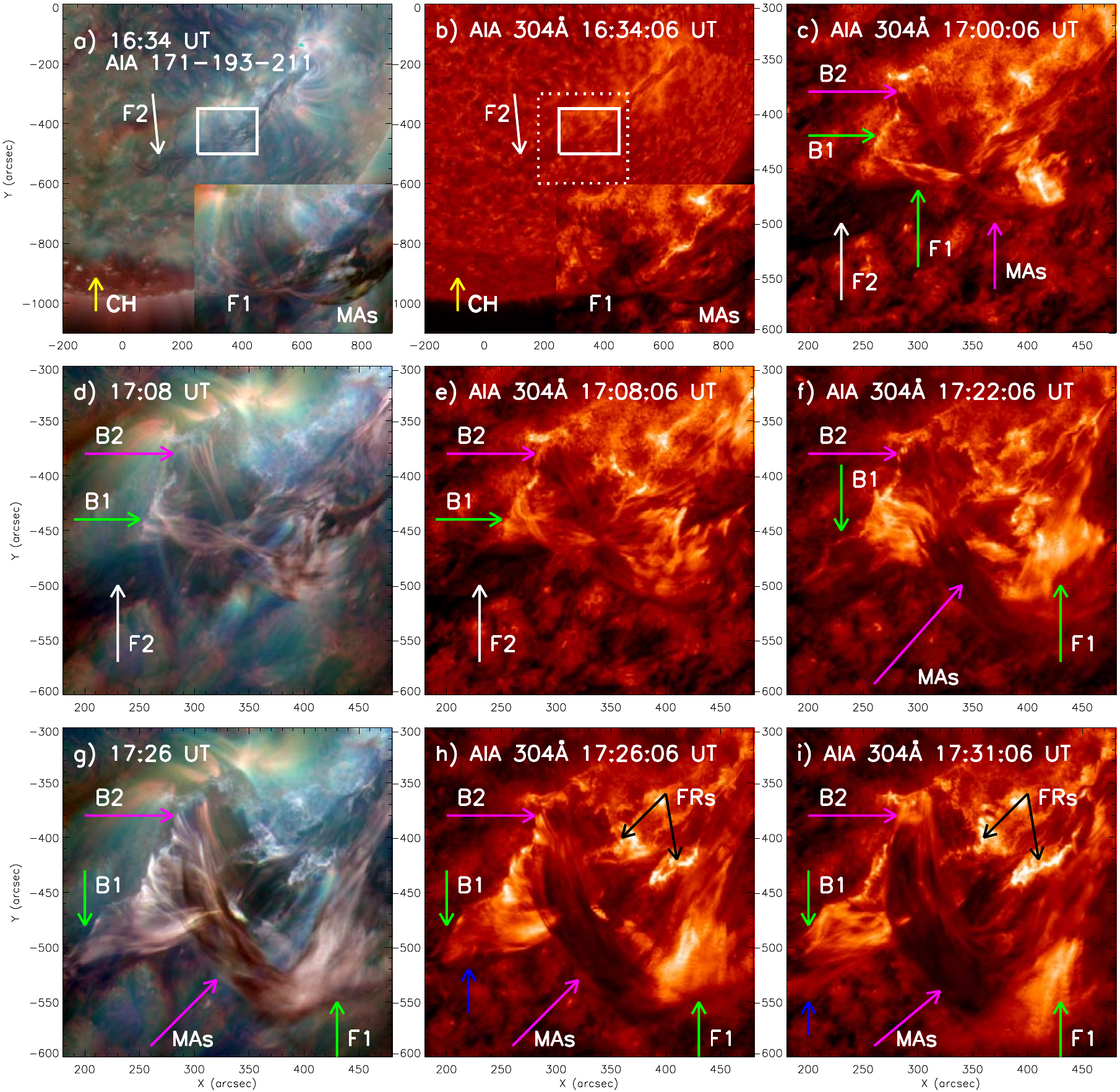}
\caption{The filament activation in composite warm passbands (left panels) and in AIA 304~{\AA} (middle and right panels). The close-up of F1 and magnetic arcades (MAs) in the solid box is shown in the right-down corner, and the dotted box indicates the FOV of panels (c)-(i). The yellow arrows show the nearby CH, and the white arrows indicate F2. The green arrows show the F1 and its east footpoints (B1), and the pink arrows indicate MAs and their east footpoints (B2). The black and blue arrows in (h)-(i) show the flare ribbons (FRs) and bright moving plasma, respectively.
\label{f2}}
\end{figure}

\clearpage

\begin{figure}
\epsscale{0.9}
\plotone{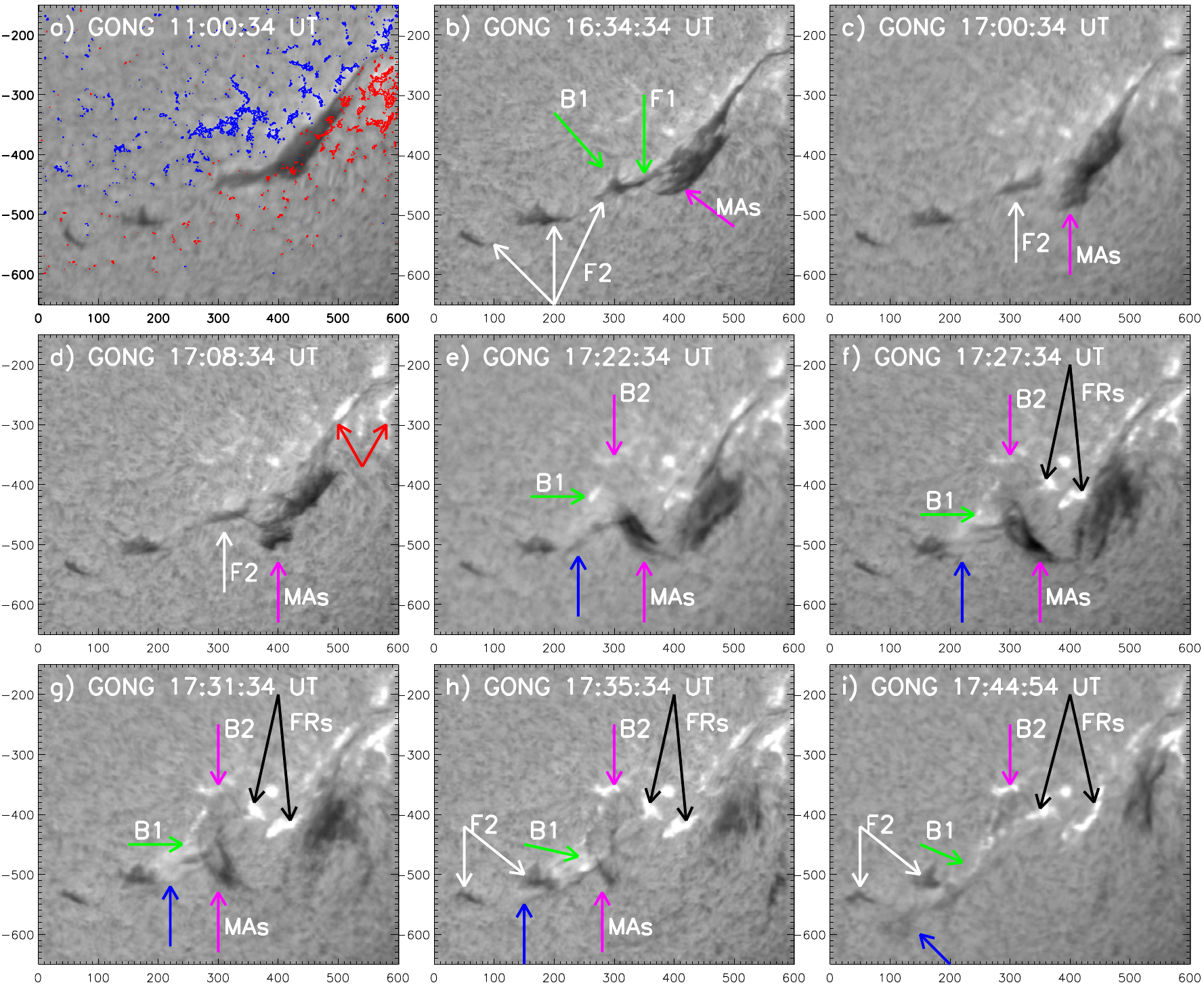}
\caption{The interaction and eruption of the filaments in GONG H$\alpha$ filtergrams. Panel (a) is superposed by the HMI magnetic field contours for the positive (red) and negative (blue) polarity, and the levels are 50, 100, and 150 Gauss. F1 and its east footpoints (B1) are shown by the green arrows, and the overlying magnetic arcades (MAs) and their east footpoints are indicated by the pink arrows. The white arrows indicate F2, and the blue arrows show the moving plasma of F2. The red arrows show the brightenings in pre-eruption phase, and the black arrows indicate the flare ribbons (FRs) after the interaction.
\label{f3}}
\end{figure}

\clearpage

\begin{figure}
\epsscale{0.9}
\plotone{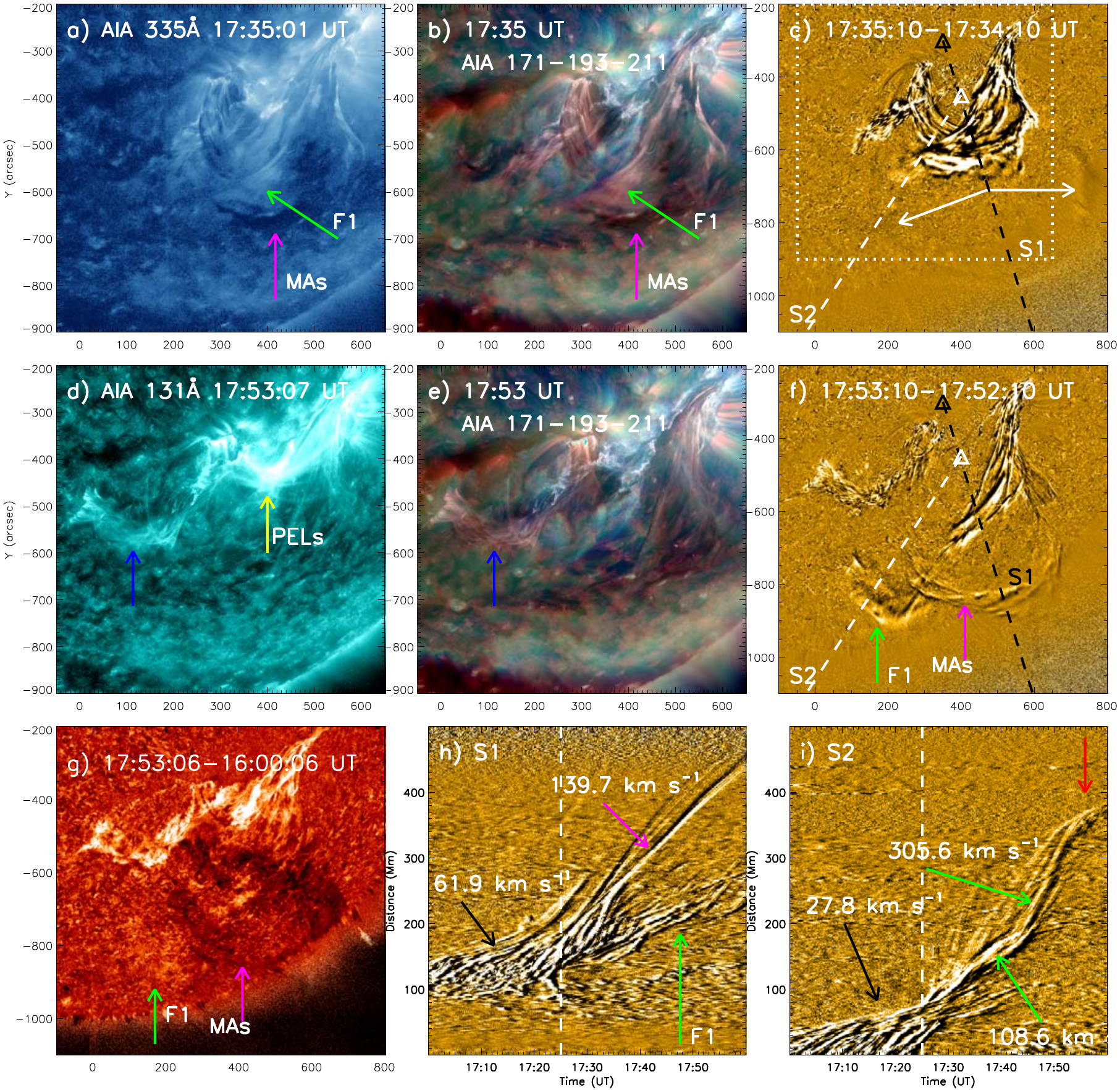}
\caption{The eruption showed in AIA images ((a) and (d)), and in composite-warm passbands ((b) and (e)), in running-difference images ((c) and (f)-(g)), and in time-slice plots ((h)-(i)) along the southward slices (S1-S2) ((c) and (f)). The green arrows show the F1, and the pink arrows indicate magnetic arcades (MAs), and the blue arrows point out the moving plasma. The white arrows mark the expanding overlying loops, and the yellow arrow indicates post-eruption loops (PELs). The dotted box in (c) outlines the FOV of (a)-(b) and (d)-(e). The black arrows show F1 and MAs in pre-eruption phase, and the red arrow points out the cease of impulsive F1 at the boundary of the nearby CH. The triangles mark the start points of S1-S2.
\label{f4}}
\end{figure}

\clearpage

\begin{figure}
\epsscale{0.9}
\plotone{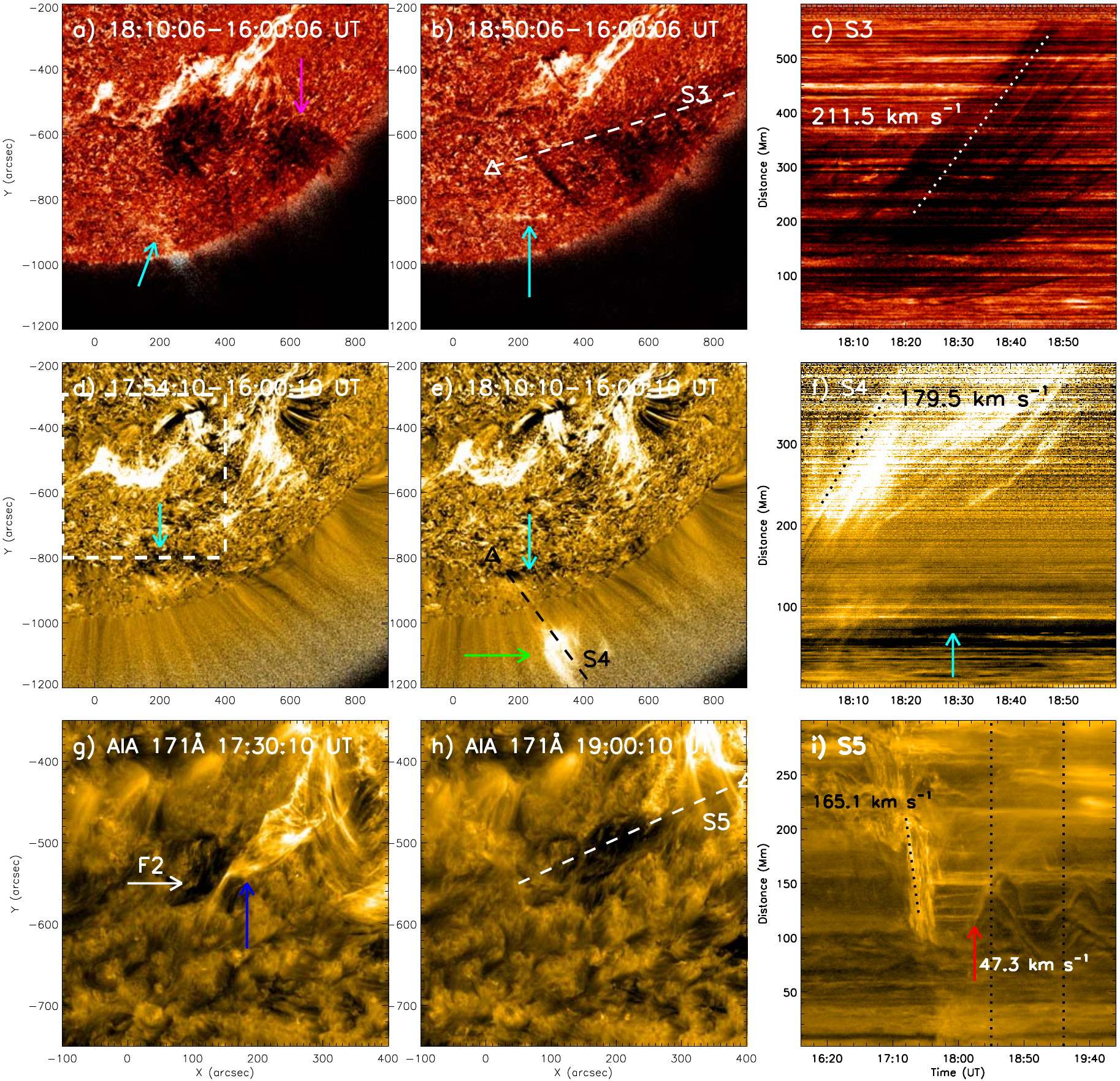}
\caption{The drift of the some plasma (pink arrow) in base-difference AIA 304~{\AA} (a-b), and in time-slice plot (c) along the northwestward slice (S3) in (b). The movement of impulsive F1 (black arrow in (e)) in base-difference AIA 171~{\AA} (d-e), and in slice-time plot (f) along the southwestward slice (S4) in (e). The cyan arrows indicate the brightenings and dimmings at the footpoints of the field lines of CH. The oscillation of F2 (white arrow) driven by moving plasma (blue arrow) in AIA 171~{\AA} (g)-(h), and in time-slice plot (i) along the southeastward slice (S5) in (h). The red arrow in (i) indicates the initial oscillation, and the vertical lines mark the times of the first two oscillation peaks to derive the period. The dashed box in (d) outlines the FOV of (g)-(h). The triangles mark the start points of S3-S5.
\label{f5}}
\end{figure}

\clearpage

\begin{figure}
\epsscale{0.9}
\plotone{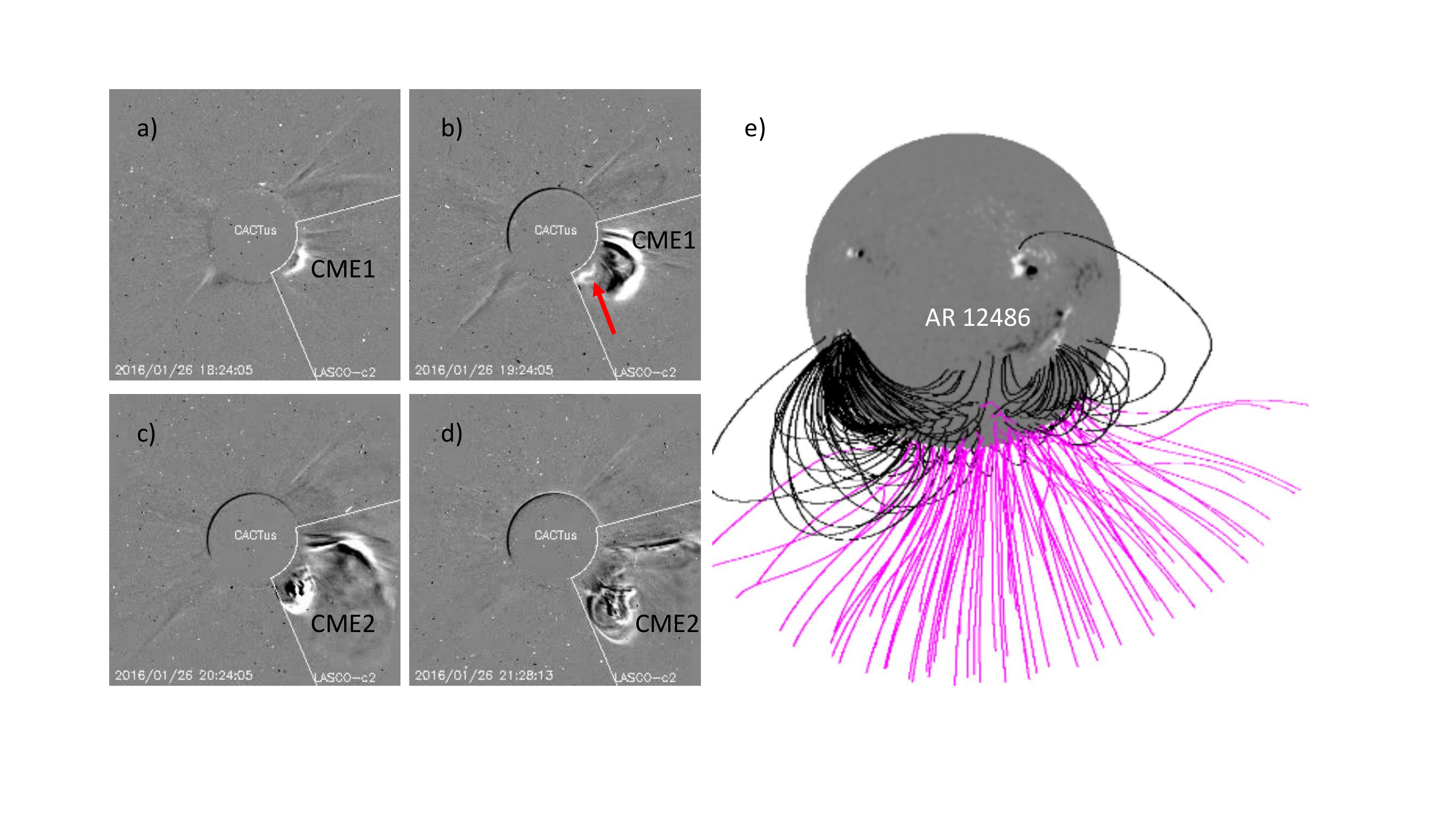}
\caption{CME evolution and PFSS extrapolation results. (a)-(d) LASCO C2 images of the CME, in which the red arrow in (b) indicates the initial front of CME2. (e) PFSS extrapolated field lines for the CH open field lines (pink) and the closed field lines (black).
\label{f6}}
\end{figure}

\clearpage

\begin{figure}
\epsscale{0.9}
\plotone{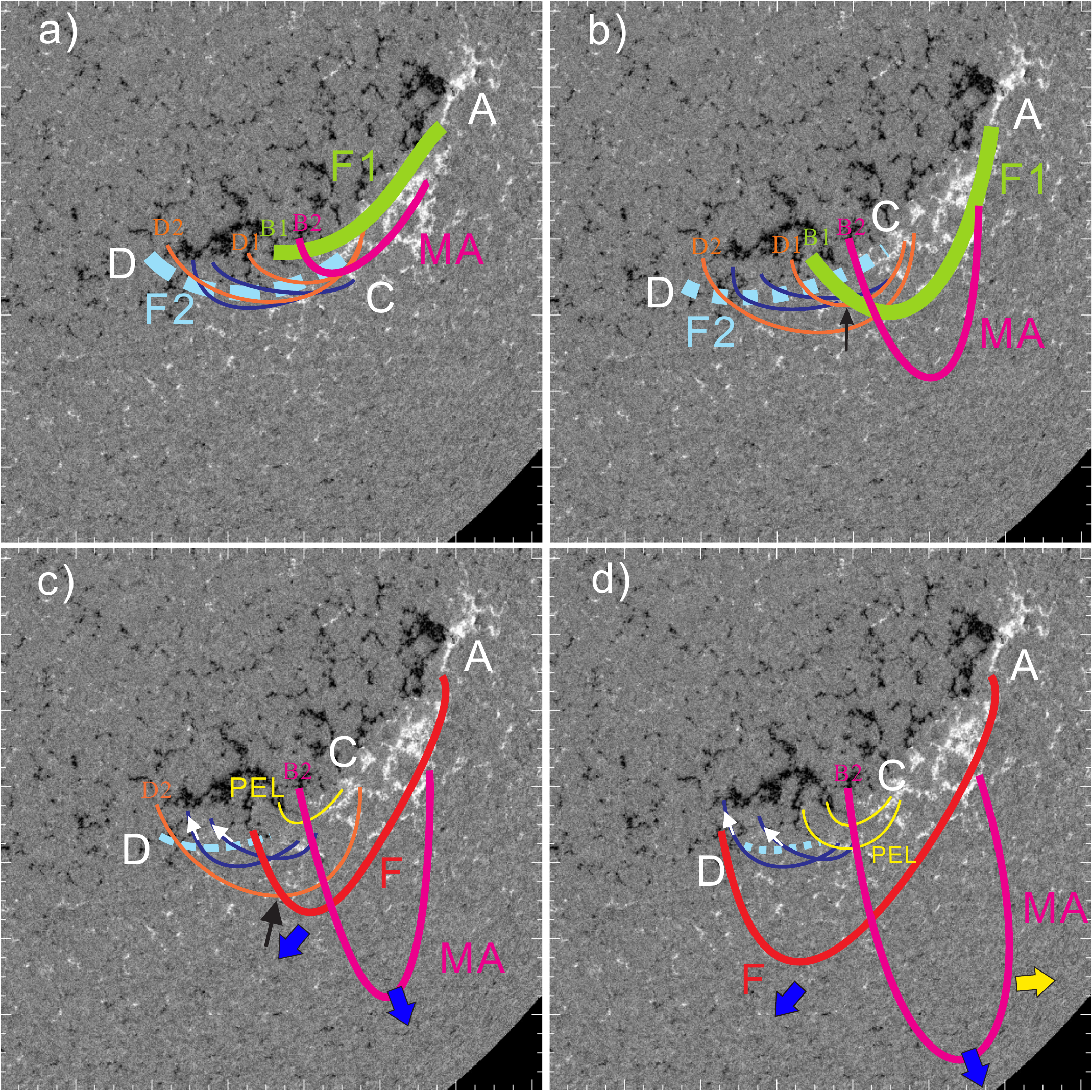}
\caption{Schematic representation of the merging reconnection during the interaction between F1 (green, AB1) and the presumed magnetic filed lines (orange) of F2 (long cyan, CD). The black arrows in (b)-(c) indicate the contact sites of two filament systems. The blue arrows show the moving directions of the newly-formed filament (red, F) and the magnetic arcades (MAs; pink), and the yellow arrow indicates the mass drift of magnetic arcades. In (c)-(d), the yellow lines show post-eruption loops (PELs), and the white arrows indicate the moving plasma along the presumed lower leftover field lines (blue lines) of the residual eastern portions (short cyan) of F2.
\label{f7}}
\end{figure}

\end{document}